\documentclass[12pt]{iopart}
\usepackage{graphicx}
\begin{document}

\title[Donut and dynamic polarization effects]{Donut and dynamic polarization effects in proton channeling through carbon nanotubes}

\author{D Borka$^1$\footnote{Corresponding author}, D J Mowbray$^2$, Z L
Mi\v{s}kovi\'{c}$^3$, S Petrovi\'{c}$^1$ and N Ne\v{s}kovi\'{c}$^1$}

\address{$^1$ Laboratory of Physics (010), Vin\v{c}a Institute of Nuclear Sciences, P.O. Box 522, 11001 Belgrade, Serbia}
\address{$^2$ Department of Physics, Center for Atomic-scale Materials Design (CAMD), Technical University of Denmark, DK-2800 Kgs. Lyngby, Denmark}
\address{$^3$ Department of Applied Mathematics, University of Waterloo, Waterloo, Ontario, Canada N2L3G1}

\ead{dusborka@vinca.rs}

\begin{abstract}
We investigate the angular and spatial distributions of protons of
the energy of 0.223 MeV after channeling through an (11,~9)
single-wall carbon nanotube of the length of 0.2 $\mu$m. The proton
incident angle is varied between 0 and 10 mrad, being close to the
critical angle for channeling. We show that, as the proton incident
angle increases and approaches the critical angle for channeling, a
ring-like structure is developed in the angular distribution - donut
effect. We demonstrate that it is the rainbow effect. When the
proton incident angle is between zero and a half of the critical
angle for channeling, the image force affects considerably the
number and positions of the maxima of the angular and spatial
distributions. However, when the proton incident angle is close to
the critical angle for channeling, its influence on the angular and
spatial distributions is reduced strongly. We demonstrate that the
increase of the proton incident angle can lead to a significant
rearrangement of the propagating protons within the nanotube. This
effect may be used to locate atomic impurities in nanotubes as well
as for creating nanosized proton beams to be used in materials
science, biology and medicine.
\end{abstract}

\submitto{\NJP}

\maketitle

\tableofcontents

\section{Introduction}

While the progress in theoretical modeling and computer simulation
of ion channeling through carbon nanotubes has reached a mature
level, as reviewed in Refs. \cite{artr05}--\cite{bell05}, the
experimental advancement in this area is still in its infancy. Since
the issues of ordering, straightening and holding nanotubes are
probably the most challenging tasks in the experimental realization
of ion channeling through them, it is not a surprise that the best
results in performing these tasks are expected when they are grown
in a dielectric medium. For example, the first experimental data on
ion channeling through nanotubes, which were reported by Zhu et al.
\cite{zhu05}, were obtained with the $He^+$ ions and array of
well-ordered multi-wall nanotubes grown in a porous anodic aluminum
oxide ($Al_2O_3$) membrane. The authors performed and compared the
results of direct measurements of the yield of ions transmitted
through the bare $Al_2O_3$ sample and the $Al_2O_3$ sample with
nanotubes.

On the other hand, the first experimental results on electron
channeling through carbon nanotubes were reported by Chai et al.
\cite{chai07}. The authors studied the transport of electrons of the
energy of 300 keV through the aligned multi-wall nanotubes of the
lengths of 0.7-3.0 $\mu$m embedded in the carbon fiber coatings. The
misalignment of the nanotubes was up to 1$^\circ$. Besides,
Berdinsky et al. \cite{berd08} succeeded in growing the single-wall
carbon nanotubes (SWCNTs) in the ion tracks etched in the $SiO_2$
layers on a Si substrate, offering an interesting possibility for
the experimental realization of ion channeling through nanotubes in
a wide range of ion energies.

Regarding the theoretical modeling and computer simulation of ion
channeling through carbon nanotubes, we note that the effect of
dynamic polarization of the nanotube atoms valence electrons by the
ion is not usually taken into account \cite{artr05}--\cite{kras05}
since its influence at very low and very high energies, of the
orders of 1 keV and 1 GeV, respectively, is negligible. However, it
is expected that at medium energies, of the order of 1 MeV, this
effect contributes significantly to the ion energy loss and gives
rise to an additional force acting on the ions, called the image
force \cite{mowb04a,mowb04b}, as it has been demonstrated in the
computer simulation of the angular distributions of protons
channeled through the SWCNTs in vacuum \cite{zhou05}. The importance
of the image force has also been emphasized in the related area of
ion transmission through cylindrical channels in metals
\cite{aris01a}--\cite{yama07} and on ions and molecules moving over
supported graphene \cite{rado07,rado08}.

When the ion channeling dynamics at very low and very high energies
is concerned, the material surrounding the carbon nanotubes serves
predominantly as their passive container. However, the ions moving
at medium energies induce the strong dynamic polarization of both
the nanotube atoms valence electrons and the surrounding material,
which in turn gives rise to a sizeable image force
\cite{mowb06,mowb07}. In these two studies, the image force was
calculated by a two-dimensional (2D) hydrodynamic model of the
nanotube atoms valence electrons while the surrounding material was
described by a frequency dependent dielectric function. On the other
hand, the image force has recently been shown to influence
significantly the rainbow effect in proton channeling through the
short SWCNTs \cite{borka06} and double-wall nanotubes in vacuum
\cite{borka07} as well as through the short SWCNTs in the dielectric
media \cite{borka08a,borka08b}. We think that it is important to
improve our understanding of the role of the image force in the
rainbow effect with nanotubes because, in analogy with the case of
surface ion channeling \cite{schu04}--\cite{wint05}, the
measurements of this effect can give precise information on both the
atomic configuration and interaction potentials within nanotubes,
which have not yet been explored completely.

However, in ion channeling experiments, the always present questions
are the ones of ion beam divergence and misalignment. So, it is
important to study the influence of the effect of dynamic
polarization of carbon nanotubes when the initial ion velocity is
not parallel to the nanotube axis. Therefore, in this paper, we
continue our investigation of the image force with the case in which
the ion incident angle is not zero. Specifically, we analyze the
angular and spatial distributions of protons of the velocity of 3
a.u. channeled through the straight (11,~9) SWCNTs of the length of
0.2 $\mu$m in vacuum. The proton incident angle is varied between 0
and 10 mrad, being close to the critical angle for channeling. This
proton velocity is chosen because the dynamic polarization effect is
the strongest in the range about it. The consideration is limited to
the case of a nanotube in vacuum because the presence of a
dielectric medium around it would introduce only a slight modifying
factor in the results of calculation \cite{borka08a,borka08b}.

It is well known that, for the ion incident angles close to the
critical angle for channeling, the donut effect develops in the
angular distributions of channeled ions. The effect was measured
with the Si and Ge crystals \cite{chad70}--\cite{ande80}, and
explained independently afterwards by the theory of crystal rainbows
\cite{nesk02,borka03}. That theory was formulated as a proper theory
of ion channeling through thin crystals \cite{petr00}, and has been
applied subsequently to ion channeling through short carbon
nanotubes \cite{petr05a}--\cite{nesk05}. It must be noted that the
donut effect has also been observed in a computer simulation of ion
propagation through nanotubes \cite{zhev98}. However, the authors
did not connect the obtained results to the rainbow effect. We
explore here the donut effect in the angular and spatial
distributions of protons channeled through a (11,~9) SWCNT in the
presence of the image force.

Regarding the angular and spatial distributions of channeled protons
to be presented in this study, corresponding to the case in which
the proton incident angle is not zero, we note that the proton
equations of motion in the transverse position plane that are solved
to generate them are 2D. This means that the case we explore is
truly 2D, unlike the cases treated in our previous studies of the
image force in carbon nanotubes, which were in fact one-dimensional
(1D) \cite{borka06}--\cite{borka08b}.

The atomic units will be used throughout the paper unless explicitly
stated otherwise.

\section{Theory}

We adopt the right Cartesian coordinate system with the $z$ axis
coinciding with the nanotube axis, the origin in the entrance plane
of the nanotube, and the $x$ and $y$ axes the vertical and
horizontal axes, respectively. The initial proton velocity,
$\vec{v}$, is taken to lie in the $yz$ plane and make angle
$\varphi$ with the $z$ axis, being the proton incident angle. The
length of the nanotube, $L$, is assumed to be large enough to allow
us to ignore the influence of the nanotube edges on the image force,
and, at the same time, small enough to neglect the energy losses of
channeled protons.

We assume that the interaction between the proton and nanotube atoms
can be treated classically using the Doyle-Turner expression
\cite{doyl68} averaged axially \cite{lind65} and azimuthally
\cite{zhev98}. This interaction is repulsive and of the short-range
character. Thus, the repulsive interaction potential in the proton
channeling through the nanotube is of the form

\begin{eqnarray}
U_\mathrm{rep}(r) = {\frac{32\pi dZ_\mathrm{1} Z_\mathrm{2} a}{3 \sqrt{3} l^2}} \nonumber\\
\times {\sum\limits_\mathrm{j = 1}^{4} {a_\mathrm{j} b_\mathrm{j}^2
I_\mathrm{0} (2a b_\mathrm{j}^{2} r) \exp \{ - b_\mathrm{j}^2}} [r^2
+ a^2]\},
\label{equ01}
\end{eqnarray}

\noindent where $Z_\mathrm{1}$ = 1 and $Z_\mathrm{2}$ = 6 are the
atomic numbers of the hydrogen and carbon atoms, respectively, $a$
is the nanotube radius, $l$ is the nanotube atoms bond length, $r =
(x^2 + y^2 )^{\frac{1}{2}}$ is the distance between the proton and
nanotube axis, $I_0$ is the modified Bessel function of the 1$^{st}$
kind and 0$^{th}$ order, and $a_\mathrm{j} = \{ 0.115, 0.188, 0.072,
0.020\}$ and $b_\mathrm{j} = \{ 0.547, 0.989, 1.982, 5.656 \}$ are
the fitting parameters (in atomic units) \cite{doyl68}.

The dynamic polarization of the nanotube by the proton is treated
via a 2D hydrodynamic model of the nanotube atoms valence electrons,
based on a jellium-like description of the ion cores making the
nanotube wall \cite{mowb04a}--\cite{mowb06}. This model includes the
axial and azimuthal averaging similar to that applied in obtaining
the corresponding repulsive interaction potential, given by Eq.
(\ref{equ01}). It finally gives the interaction potential between
the proton and its image, $U_\mathrm{im}(r,t)$, which is stationary
in the coordinate system moving with the proton and depends on its
velocity. This interaction is attractive and of the long-range
character. The details of derivation of the expression for
$U_\mathrm{im}(r,t)$ are given elsewhere
\cite{mowb04a}--\cite{borka08a}. Consequently, the total interaction
potential in the proton channeling through the nanotube is

\begin{equation}
U(r,t) = U_\mathrm{rep}(r)+U_\mathrm{im}(r,t).
\label{equ02}
\end{equation}

The proton equations of motion we solve are

\begin{equation}
m\ddot x(t) = - \frac{{\partial U(r,t)}}{\partial x},
\label{equ03}
\end{equation}

\begin{equation}
m\ddot y(t) = - \frac{{\partial U(r,t)}}{\partial y},
\label{equ04}
\end{equation}

\noindent where $m$ is the proton mass. They are subject to the
initial conditions for the transverse components of the proton
velocity that are

\begin{equation}
\dot x(t = 0) = 0,
\label{equ05}
\end{equation}

\begin{equation}
\dot y(t = 0) = v\sin \varphi  \approx v\varphi.
\label{equ06}
\end{equation}

The longitudinal proton motion is treated as uniform with the
initial condition for the longitudinal component of the proton
velocity that is $\dot z(t = 0) = v\cos \varphi  \approx v$. As a
result, the longitudinal component of the proton position is $z(t) =
vt$. Equations (5) and (6) are solved numerically. The angular and
spatial distributions of transmitted protons are generated using a
Monte Carlo computer simulation code. The components of the proton
impact parameter, $x_0$ and $y_0$, are chosen randomly from a
uniform distribution within the cross-sectional area of the nanotube
and its entrance plane. With $l$ = 0.144 nm \cite{sait01}, we obtain
that $a$ = 0.689 nm. If the proton impact parameter falls inside
annular interval $[a - a_{sc},a]$, where $a_{sc}  = [9\pi^2 /(128
Z_2 )]^{\frac{1}{3}} a_0$ is the screening radius and $a_0$ the Bohr
radius, the proton is treated as if it is backscattered and is
disregarded. The initial number of protons is about 1~000~000.

The components of the proton scattering angle, $\Theta_x$ and
$\Theta_y$, are obtained via expressions $\Theta_x  = V_x / v$ and
$\Theta_y = V_y / v$, where $V_x$ and $V_y$ are the final transverse
components of the proton velocity, which are obtained, together with
the final transverse components of the proton position, $X$ and $Y$,
as the solutions of Eqs. (\ref{equ05}) and (\ref{equ06}). The proton
channeling through the nanotube can be analyzed via the mapping of
the impact parameter plane, the $x_0 y_0$ plane, to the scattering
angle plane, the $\Theta_x \Theta_y$ plane \cite{petr00}. The
corresponding total interaction potential, given by Eq.
(\ref{equ02}), is axially symmetric. This means that, if the initial
proton velocities were parallel to the nanotube axis, this mapping
would be 1D. However, the initial proton velocities are not parallel
to the nanotube axis, and the mapping is 2D. Since the proton
scattering angle is small, its differential transmission cross
section is given by

\begin{equation}
\sigma  =  - \frac{1}{\left| J_\Theta  \right|},
\label{equ07}
\end{equation}

\noindent where $J_{\Theta}$ is the Jacobian of the mapping,

\begin{equation}
J_\Theta = \frac{\partial \Theta_x}{\partial x_0}\frac{\partial
\Theta_y}{\partial y_0} - \frac{\partial \Theta_x}{\partial
y_0}\frac{\partial \Theta_y}{\partial x_0}.
\label{equ08}
\end{equation}

Thus, equation $J_{\Theta}$ = 0 determines the lines in the impact
parameter plane along which the proton differential transmission
cross section is singular. The images of these lines in the
scattering angle plane are the rainbow lines in this plane
\cite{petr00}.

We can analyze in a similar way the mapping of the impact parameter
plane (the $x_0 y_0$ plane), which is the entrance plane of the
nanotube and the initial transverse position plane, to the exit
plane of the nanotube or the final transverse position plane, the
plane. The Jacobian of this mapping is

\begin{equation}
J_R = \frac{\partial X}{\partial x_0}\frac{\partial Y}{\partial y_0}
- \frac{\partial X}{\partial y_0}\frac{\partial Y}{\partial x_0}.
\label{equ09}
\end{equation}

The rainbow lines in the final transverse position plane are the
images of the lines in the impact parameter plane determined by
equation $J_R$ = 0.

\section{Results and discussion}

\begin{figure}
\centering
\includegraphics[width=0.45\textwidth]{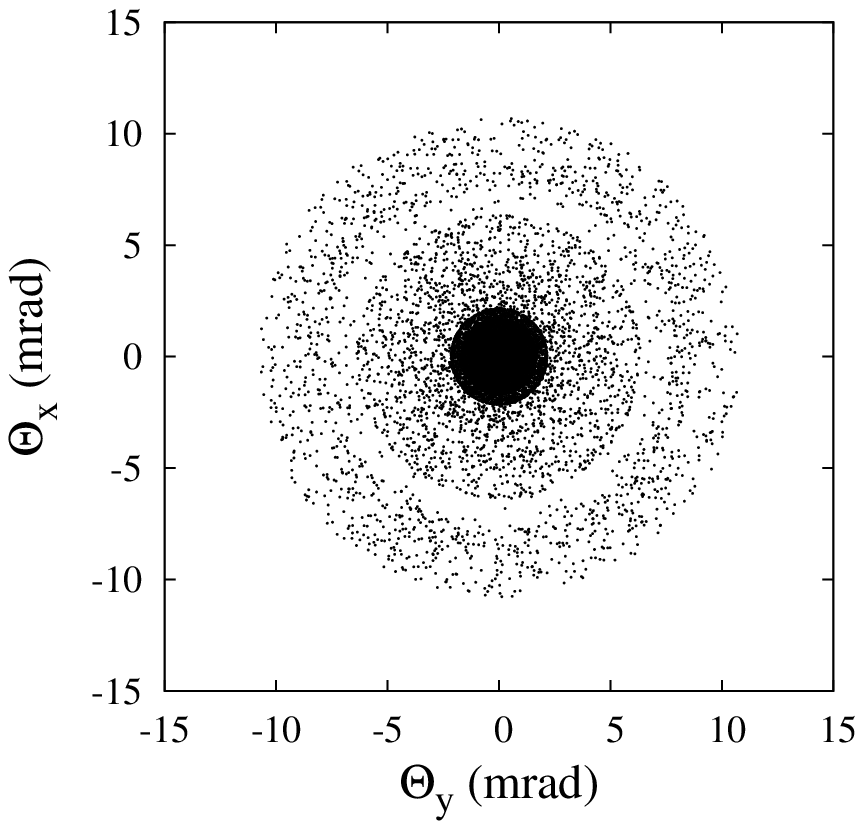}
\caption{The angular distribution of protons channeled in the
(11,~9) SWCNT with the inclusion of the image force when the proton
incident angle $\varphi$ = 0. The proton velocity is $v$ = 3 a.u.,
the nanotube radius $a$ = 0.689 nm, and the nanotube length $L$ =
0.2 $\mu$m.}
\label{fig01}
\end{figure}

\begin{figure}
\centering
\includegraphics[width=0.45\textwidth]{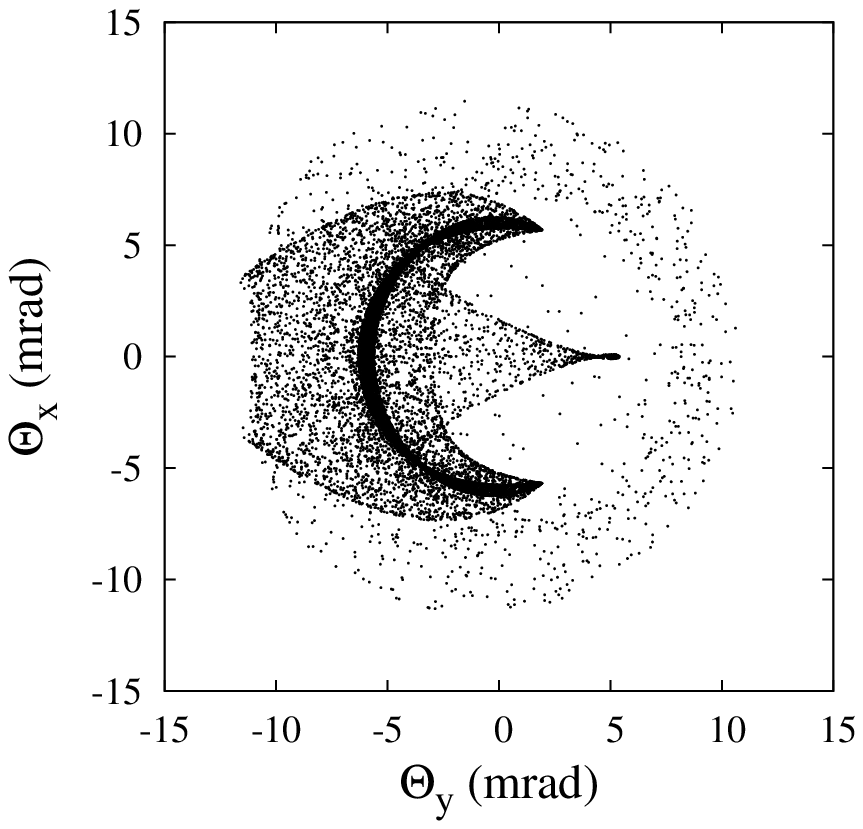}
\caption{The angular distribution of protons channeled in the
(11,~9) SWCNT with the inclusion of the image force when the proton
incident angle $\varphi$ = 6 mrad. The proton velocity is $v$ = 3
a.u., the nanotube radius $a$ = 0.689 nm, and the nanotube length
$L$ = 0.2 $\mu$m.}
\label{fig02}
\end{figure}

\begin{figure}
\centering
\includegraphics[width=0.45\textwidth]{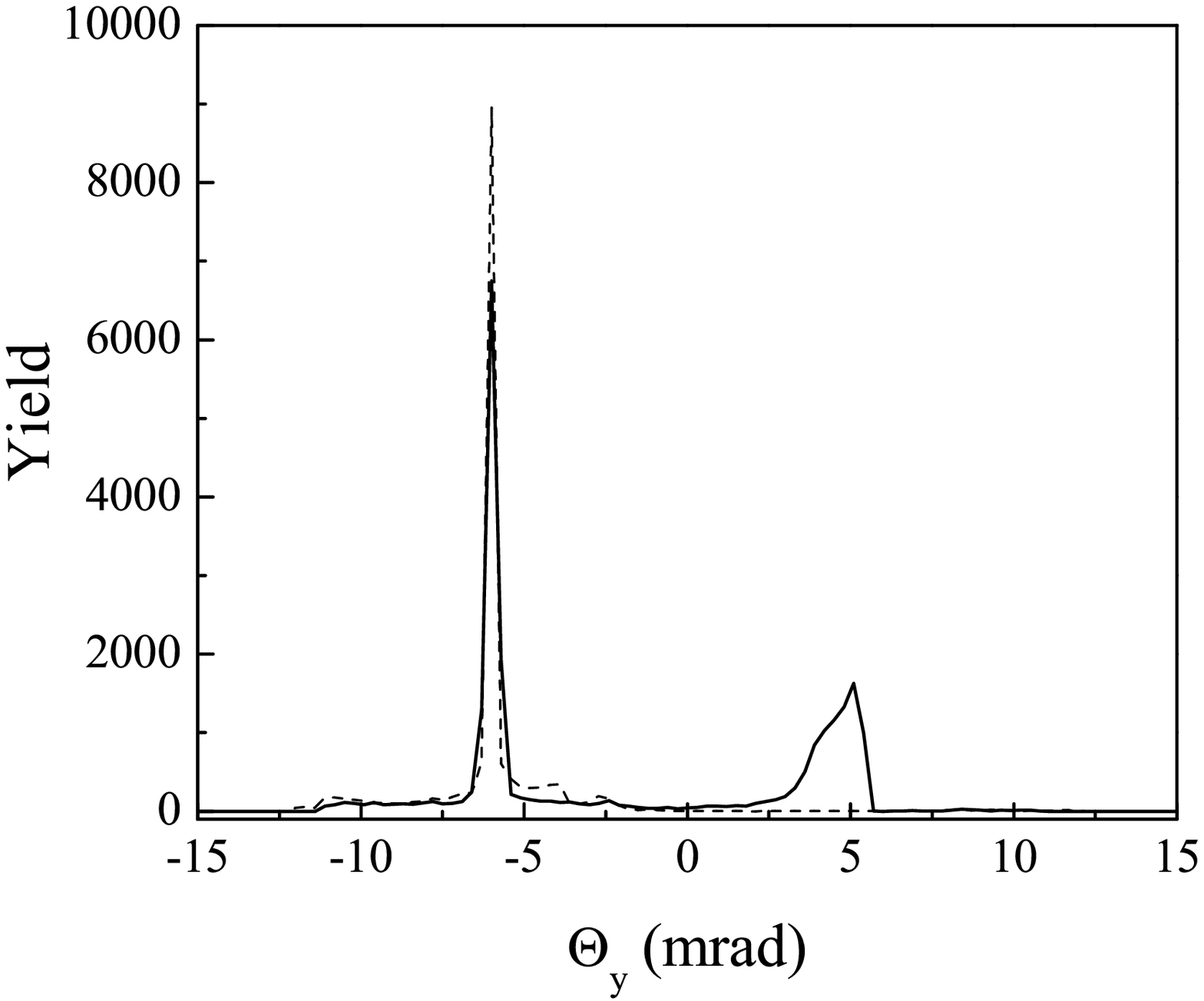}
\caption{The distribution along the $\Theta_y$ axis of protons
channeled in the (11,~9) SWCNT with the image force taken into
account -- the solid curve -- and without it -- the dashed curve --
when the proton incident angle $\varphi$ = 6 mrad. The proton
velocity is $v$ = 3 a.u., the nanotube radius $a$ = 0.689 nm, and
the nanotube length $L$ = 0.2 $\mu$m. The former curve corresponds
to the angular distribution shown in Fig. 2.}
\label{fig03}
\end{figure}

\begin{figure}
\centering
\includegraphics[width=0.45\textwidth]{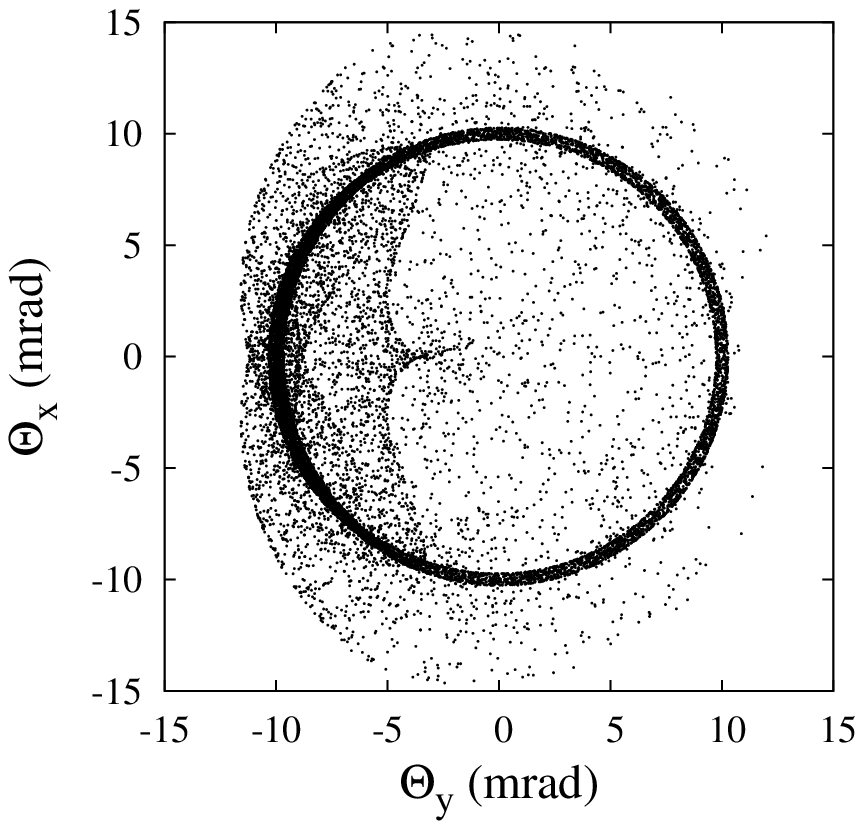}
\caption{The angular distribution of protons channeled in the
(11,~9) SWCNT with the inclusion of the image force when the proton
incident angle $\varphi$ = 10 mrad. The proton velocity is $v$ = 3
a.u., the nanotube radius $a$ = 0.689 nm, and the nanotube length
$L$ = 0.2 $\mu$m.}
\label{fig04}
\end{figure}

\begin{figure}
\centering
\includegraphics[width=0.45\textwidth]{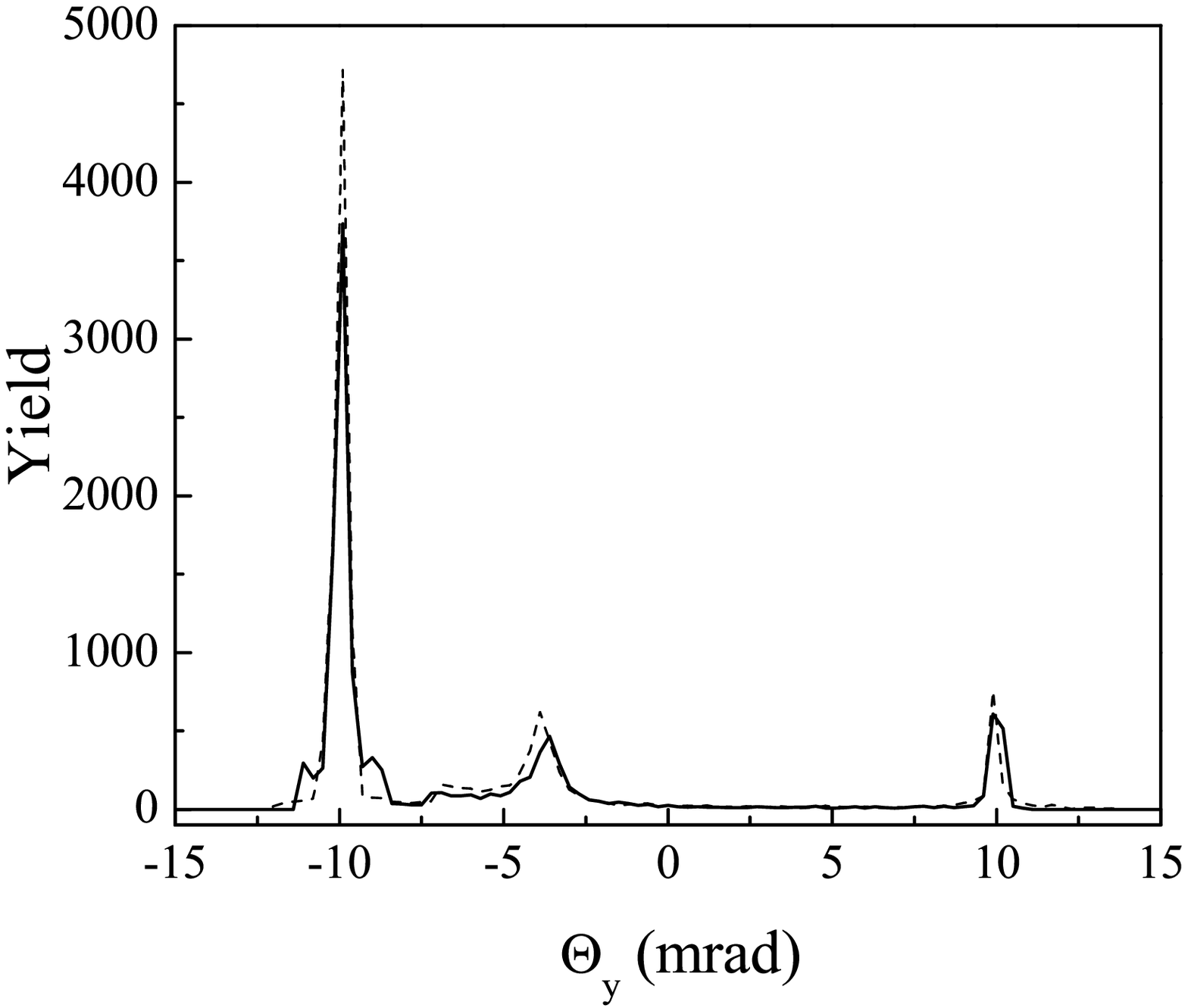}
\caption{2The distribution along the $\Theta_y$ axis of protons
channeled in the (11,~9) SWCNT with the image force taken into
account -- the solid curve -- and without it -- the dashed curve --
when the proton incident angle $\varphi$ = 10 mrad. The proton
velocity is $v$ = 3 a.u., the nanotube radius $a$ = 0.689 nm, and
the nanotube length $L$ = 0.2 $\mu$m. The former curve corresponds
to the angular distribution shown in Fig. 4.}
\label{fig05}
\end{figure}

\begin{figure}
\centering
\includegraphics[width=0.45\textwidth]{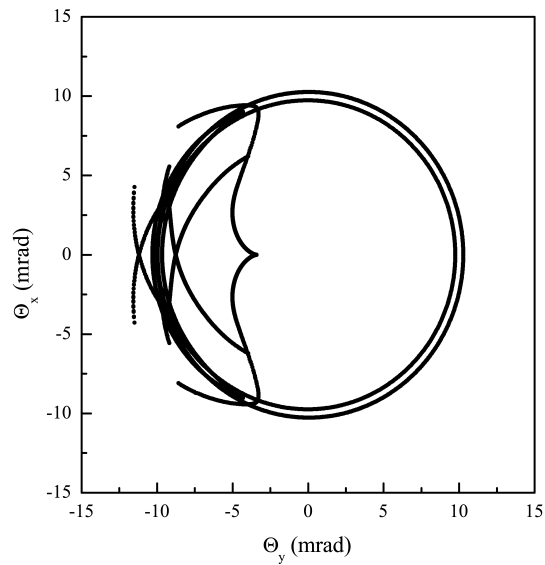}
\caption{The rainbow lines in the scattering angle plane for the
protons channeled in the (11,~9) SWCNT with the inclusion of the
image force when the proton incident angle $\varphi$ = 10 mrad. The
proton velocity is $v$ = 3 a.u., the nanotube radius $a$ = 0.689 nm,
and the nanotube length $L$ = 0.2 $\mu$m.}
\label{fig06}
\end{figure}

\begin{figure}
\centering
\includegraphics[width=0.45\textwidth]{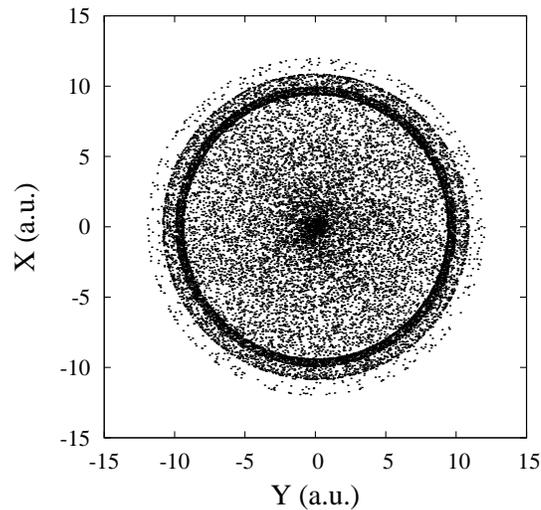}
\caption{The spatial distribution of protons channeled in the
(11,~9) SWCNT with the image force included when the proton incident
angle $\varphi$ = 0. The proton velocity is $v$ = 3 a.u., the
nanotube radius $a$ = 0.689 nm, and the nanotube length $L$ = 0.2
$\mu$m.}
\label{fig07}
\end{figure}

\begin{figure}
\centering
\includegraphics[width=0.40\textwidth]{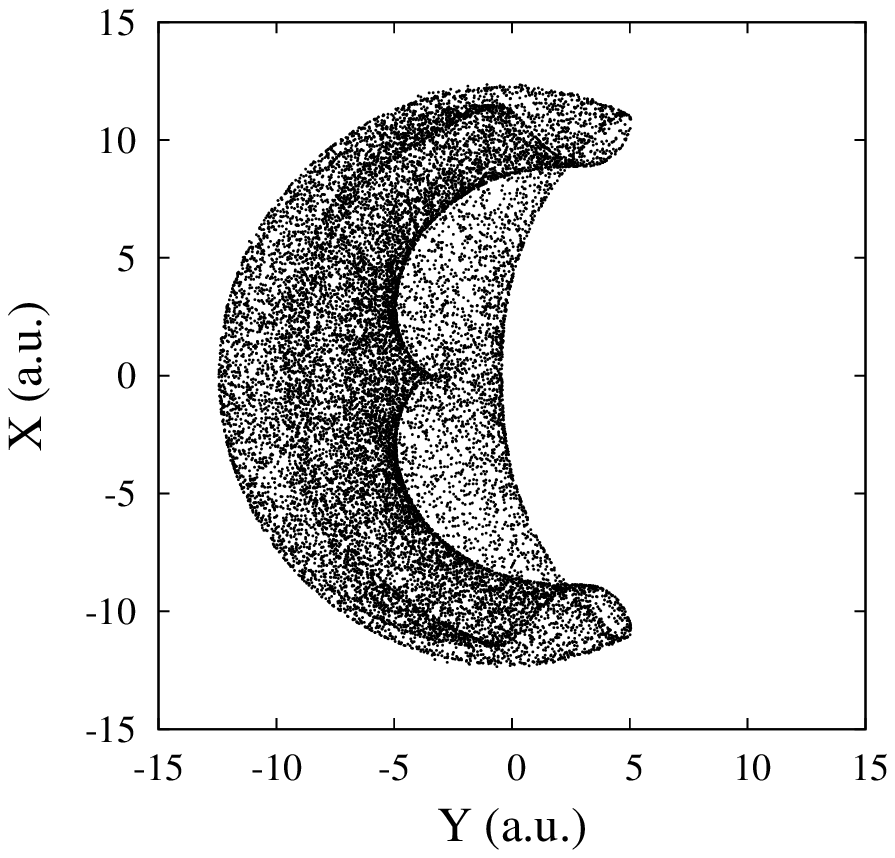}
\caption{The spatial distribution of protons channeled in the
(11,~9) SWCNT with the inclusion of the image force when the proton
incident angle $\varphi$ = 10 mrad. The proton velocity is $v$ = 3
a.u., the nanotube radius $a$ = 0.689 nm, and the nanotube length
$L$ = 0.2 $\mu$m.}
\label{fig08}
\end{figure}

\begin{figure}
\centering
\includegraphics[width=0.45\textwidth]{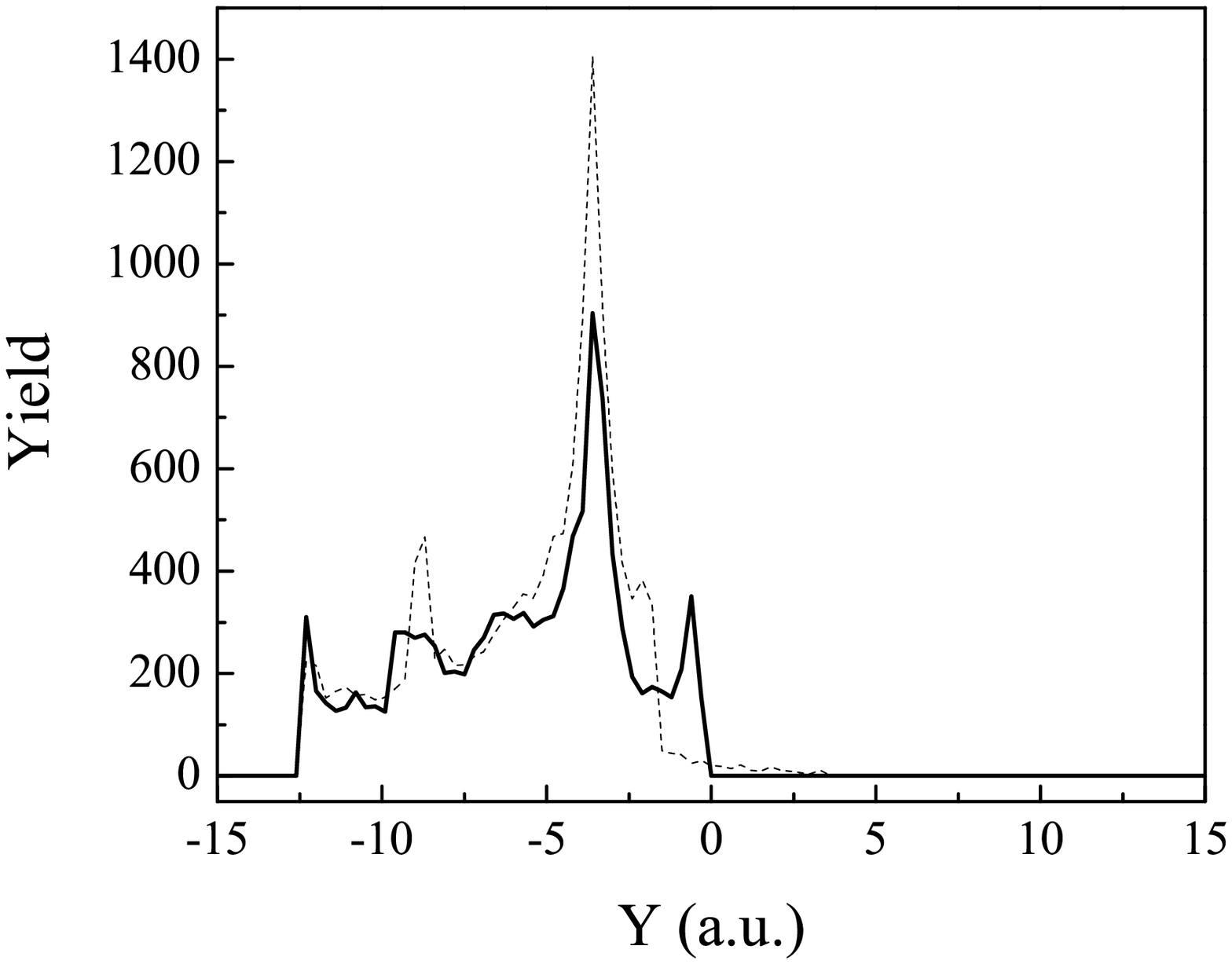}
\caption{The distribution along the $Y$ axis of protons channeled in
the (11,~9) SWCNT with the image force taken into account -- the
solid curve -- and without it -- the dashed curve -- when the proton
incident angle $\varphi$ = 10 mrad. The proton velocity is $v$ = 3
a.u., the nanotube radius $a$ = 0.689 nm, and the nanotube length
$L$ = 0.2 $\mu$m. The former curve corresponds to the spatial
distribution given in Fig. 8.}
\label{fig09}
\end{figure}

\begin{figure}
\centering
\includegraphics[width=0.40\textwidth]{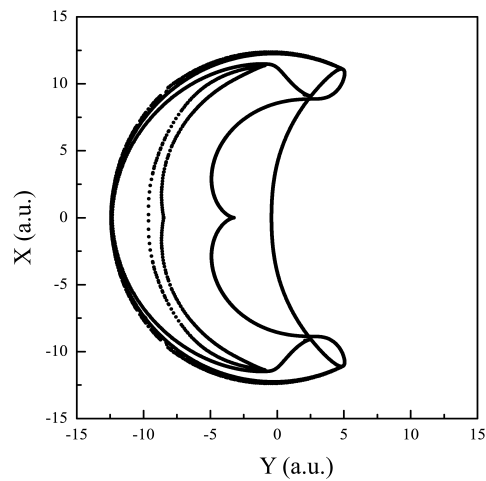}
\caption{The rainbow lines in the final transverse position plane
for the protons channeled in the (11,~9) SWCNT with the image force
included when the proton incident angle $\varphi$ = 10 mrad. The
proton velocity is $v$ = 3 a.u., the nanotube radius $a$ = 0.689 nm,
and the nanotube length $L$ = 0.2 $\mu$m.}
\label{fig10}
\end{figure}

\begin{figure}
\centering
\includegraphics[width=0.45\textwidth]{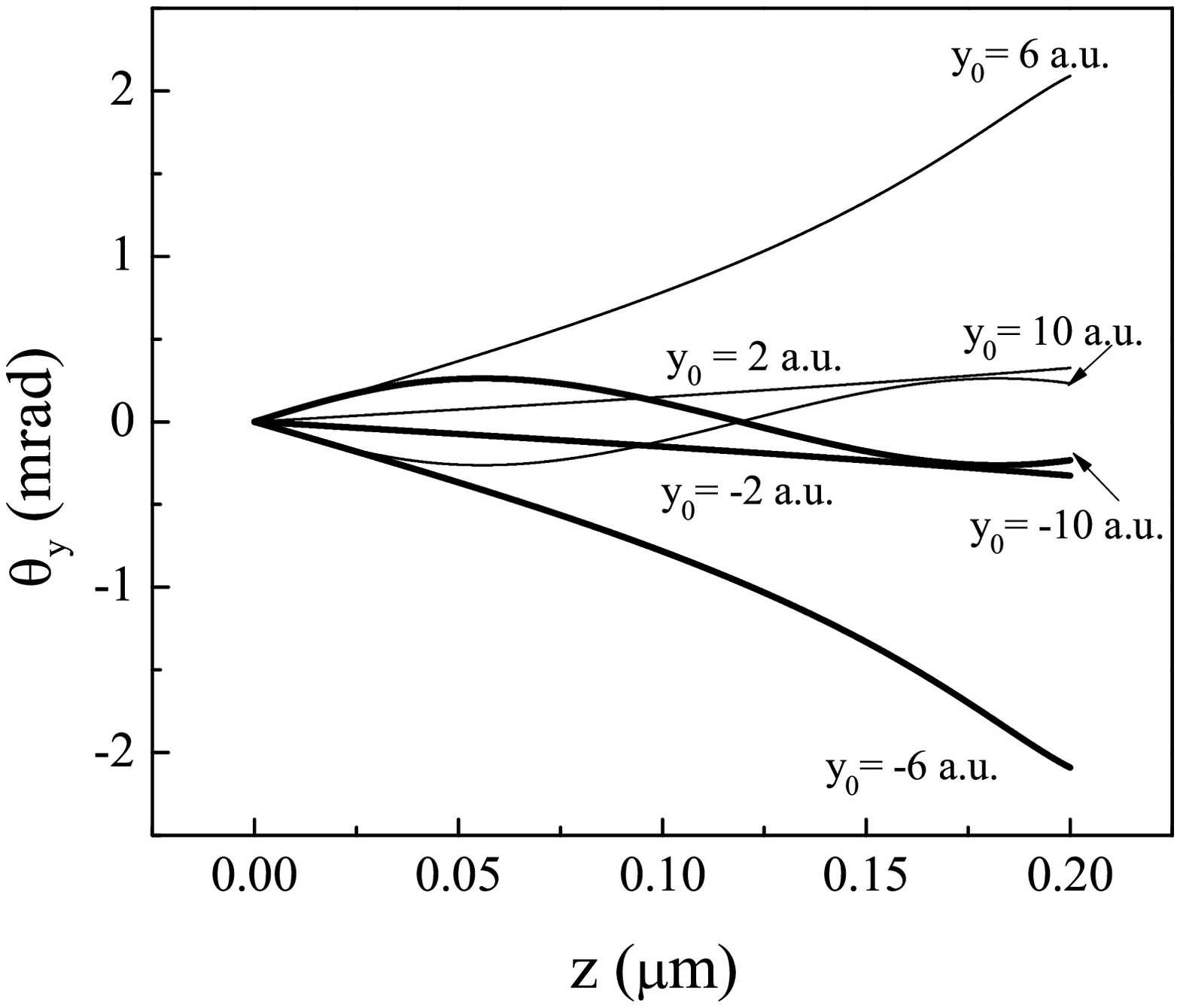}
\caption{The dependence of the $y$ component of the proton
scattering angle on the $z$ component of its position in the
channeling through the (11,~9) SWCNT with the inclusion of the image
force when the proton incident angle $\varphi$ = 0 for the $x$
component of the proton impact parameter $x_0$ = 0 and the $y$
component of its impact parameter $y_0$ = $\pm2$, $\pm6$ and $\pm10$
a.u. The proton velocity is $v$ = 3 a.u., the nanotube radius $a$ =
0.689 nm, and the nanotube length $L$ = 0.2 $\mu$m.}
\label{fig11}
\end{figure}

\begin{figure}
\centering
\includegraphics[width=0.45\textwidth]{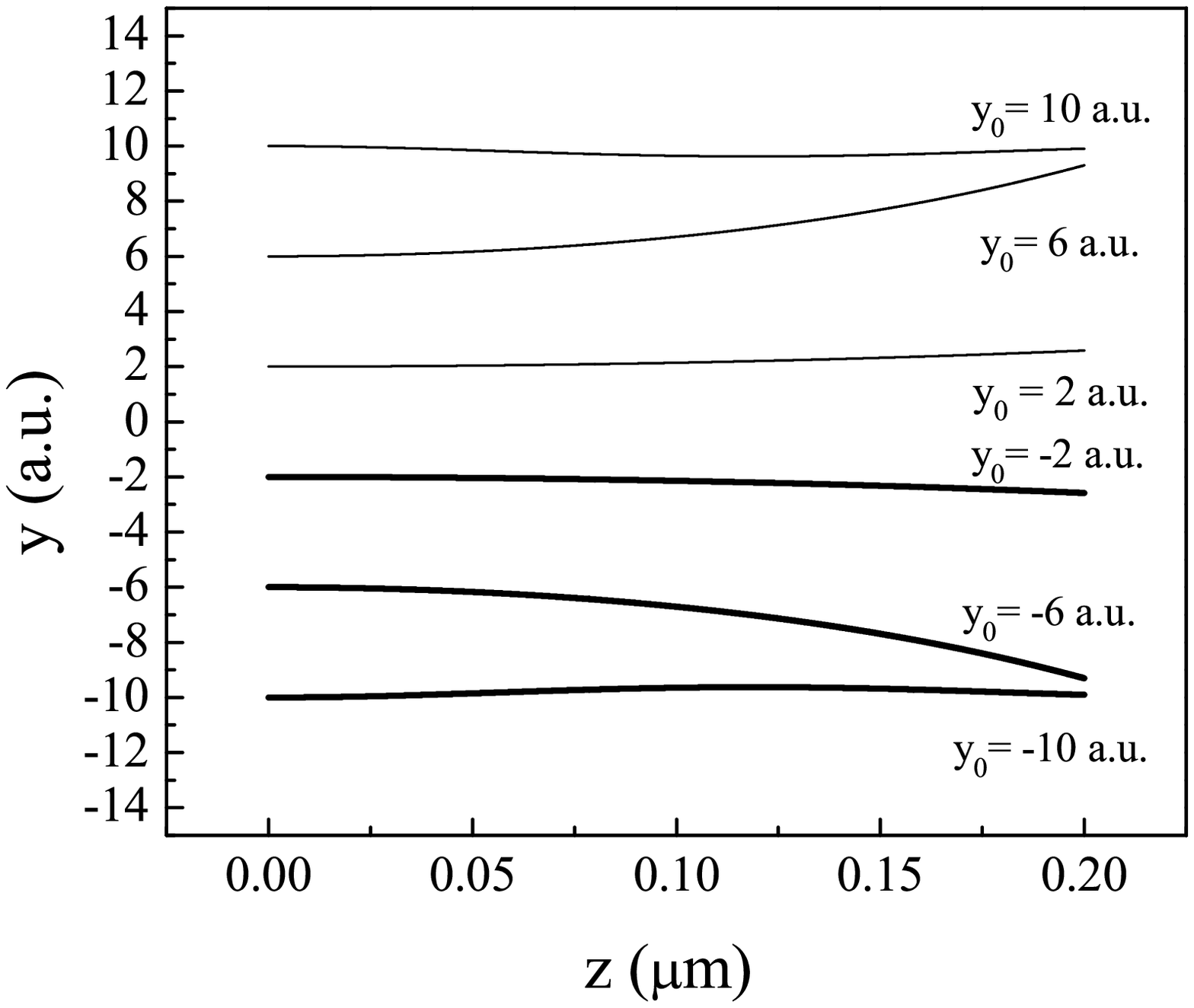}
\caption{The dependence of the $y$ component of the proton position
on the $z$ component of its position in the channeling through the
(11,~9) SWCNT with the inclusion of the image force when the proton
incident angle $\varphi$ = 0 for the $x$ component of the proton
impact parameter $x_0$ = 0 and the $y$ component of its impact
parameter $y_0$ = $\pm2$, $\pm6$ and $\pm10$ a.u. The proton
velocity is $v$ = 3 a.u., the nanotube radius   = 0.689 nm, and the
nanotube length $L$ = 0.2 $\mu$m.}
\label{fig12}
\end{figure}

\begin{figure}
\centering
\includegraphics[width=0.45\textwidth]{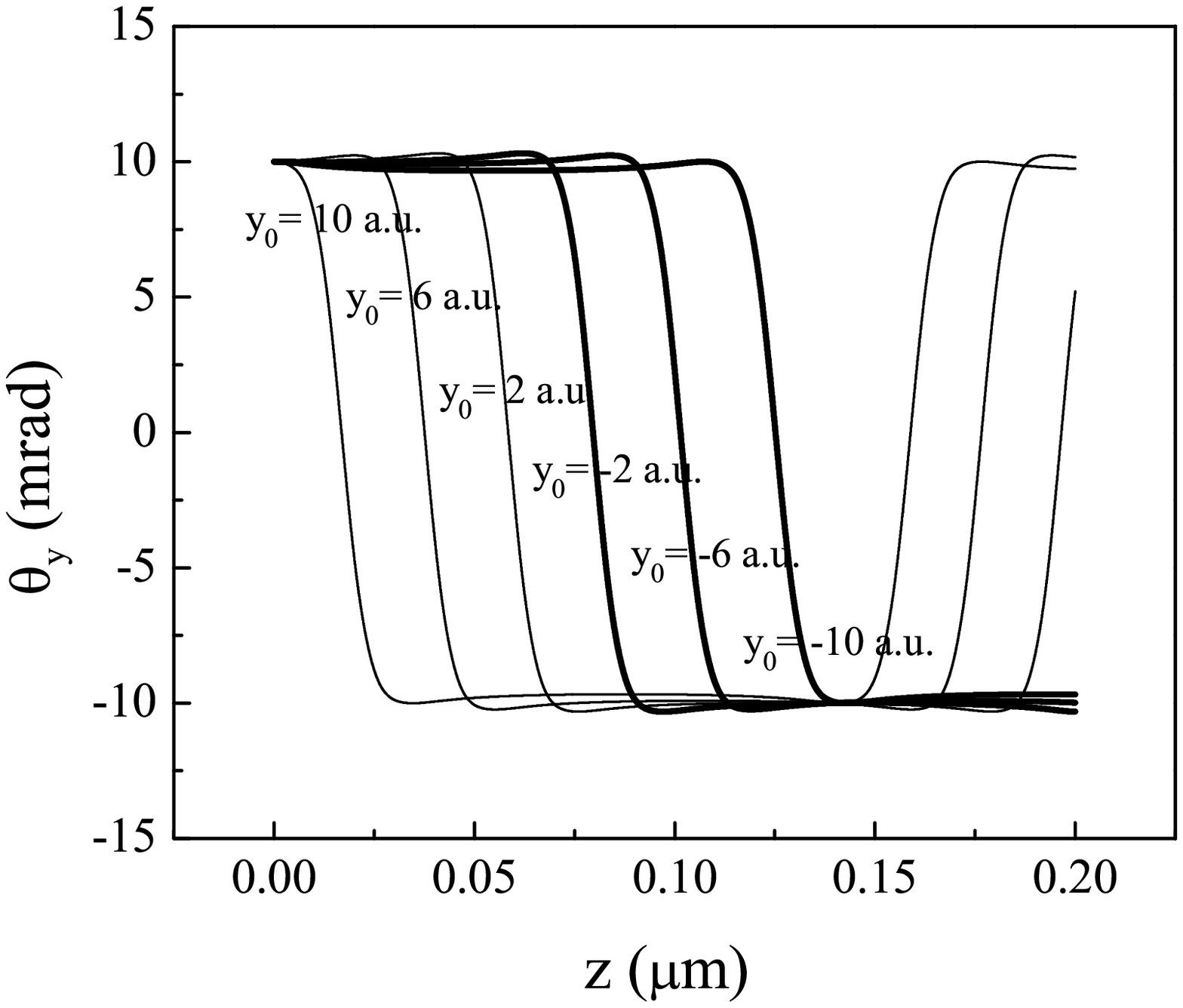}
\caption{The dependence of the $y$ component of the proton
scattering angle on the $z$ component of its position in the
channeling through the (11,~9) SWCNT with the inclusion of the image
force when the proton incident angle $\varphi$ = 10 mrad for the $x$
component of the proton impact parameter $x_0$ = 0 and the $y$
component of its impact parameter $y_0$ = $\pm2$, $\pm6$ and $\pm10$
a.u. The proton velocity is $v$ = 3 a.u., the nanotube radius $a$ =
0.689 nm, and the nanotube length $L$ = 0.2 $\mu$m.}
\label{fig13}
\end{figure}

\begin{figure}
\centering
\includegraphics[width=0.45\textwidth]{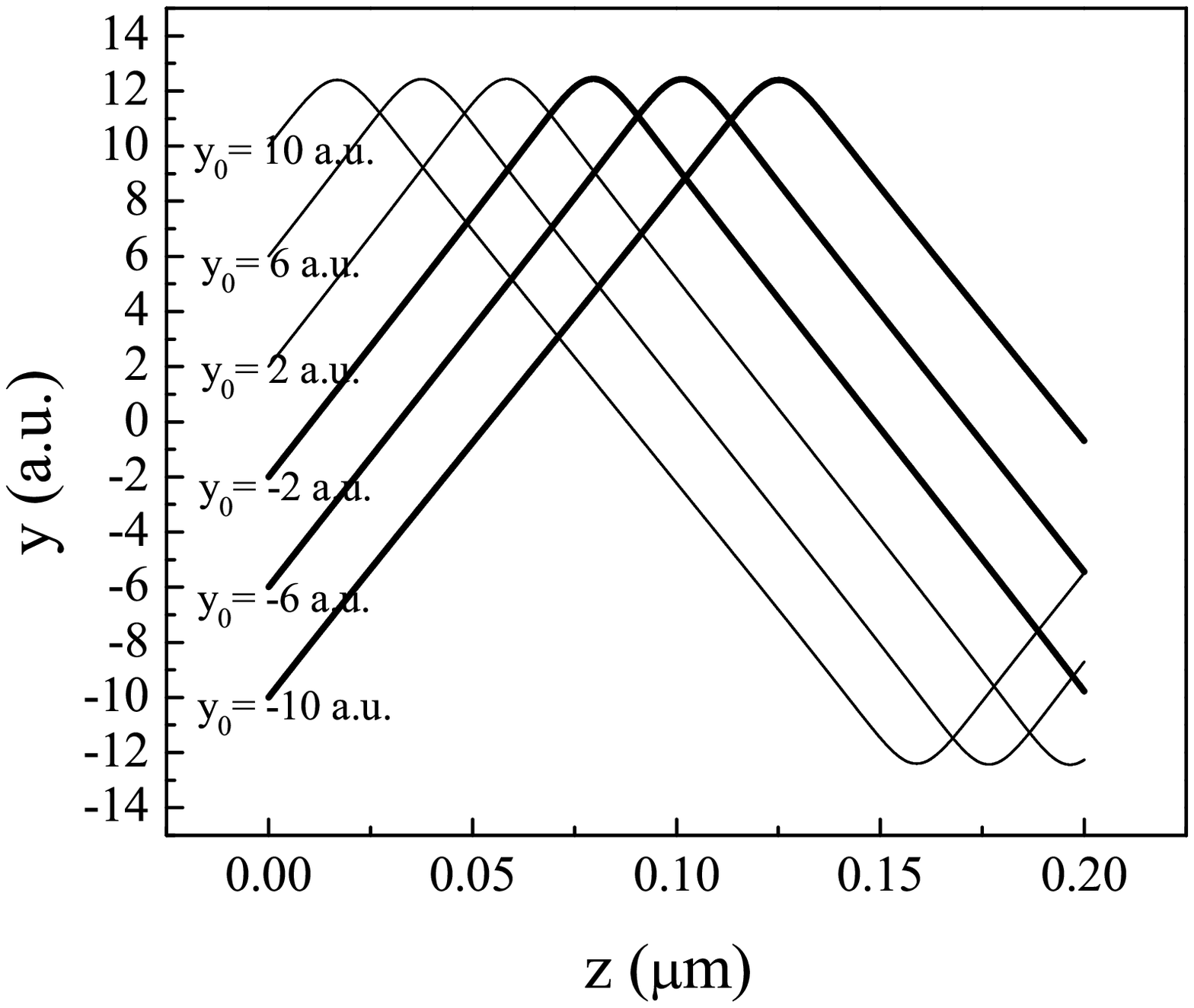}
\caption{The dependence of the $y$ component of the proton position
on the $z$ component of its position in the channeling through the
(11,~9) SWCNT with the image force taken into account when the
proton incident angle $\varphi$ = 10 mrad for the $x$ component of
the proton impact parameter $x_0$ = 0 and the $y$ component of its
impact parameter $y_0$ = $\pm2$, $\pm6$ and $\pm10$ a.u. The proton
velocity is $v$ = 3 a.u., the nanotube radius $a$ = 0.689 nm, and
the nanotube length $L$ = 0.2 $\mu$m.}
\label{fig14}
\end{figure}

Let us now analyze the angular and spatial distributions of protons
channeled in the (11,~9) SWCNT of the length of 0.2 $\mu$m. In all
the cases to be studied, the initial proton velocity will be $v$ = 3
a.u., corresponding to the initial proton energy of 0.223 MeV, while
the incident proton angle, $\varphi$, will be varied between 0 and
10 mrad. The maximal proton incident angle will be close to the
critical angle for channeling, $\psi_c$, being about 11 mrad. The
analysis will also include the typical proton trajectories through
the nanotube in the proton phase space.

In Figs. \ref{fig01}-\ref{fig06} we shall display the evolution of
the angular distribution of channeled protons with the increase of
$\varphi$. In particular, we shall analyze the development of a
ring-like structure in the angular distribution under the influence
of the image force.

The scatter plot shown in Fig. \ref{fig01} represents the angular
distribution of channeled protons for $\varphi$ = 0 with the image
force included. The corresponding angular distribution without the
inclusion of the image force contains in its central part only a
maximum at the origin. This means that the non-monotonic character
of the central part of the angular distribution with the image force
included is due to the effect of dynamic polarization. This was
discussed in one of our previous papers \cite{borka06}. We presented
in it the distributions of channeled protons along the $\Theta_y$
axis (in the scattering angle plane) with and without the image
force included, which were in fact (for $\varphi$ = 0) the radial
yields of channeled protons. The conclusion of the discussion was
that the maxima of the radial yield, appearing when the image force
was included, were due to the rainbow effect.

Fig. \ref{fig02} gives the angular distribution of channeled protons
for $\varphi$ = 6 mrad with the image force included. One can notice
easily about a half of a ring-like structure, with an exceptionally
high yield of channeled protons. This is the precursor of the effect
known as the donut effect, which is connected to the misalignment of
the proton beam and nanotube axis. In addition, the angular
distribution contains several intricately shaped regions with lower
yields of channeled protons. We show in Fig. \ref{fig03} the
corresponding distribution of channeled protons along the $\Theta_y$
axis with and without the image force included. The sharp maximum of
this distribution, appearing at -6.0 mrad, is due to the donut
effect. It is evident that the image force makes this maximum
weaker. On the other hand, the origin of the broad maximum of the
distribution, located at -5.1 mrad, is solely the image force. Thus,
we can conclude that for the median values of $\varphi$, between 0
and about $\psi_c / 2$, the image force still plays a significant
role in generating the angular distribution.

We show in Fig. \ref{fig04} the angular distribution of channeled
protons for $\varphi$ = 10 mrad with the image force included. One
can see clearly the whole ring-like structure, with an exceptionally
high yield of channeled protons. This is the fully developed donut
effect. As it has been already said, the corresponding value of
$\varphi$ is close to the value of $\psi_c$. In addition, as in Fig.
\ref{fig02}, the angular distribution contains several intricately
shaped regions with lower but distinctly graded yields of channeled
protons, with the very clear boundaries between them. Fig.
\ref{fig05} gives the corresponding distribution of channeled
protons along the $\Theta_y$ axis with and without the image force
included. It is evident that, when $\varphi$ is close to $\psi_c$,
the role of the effect of dynamic polarization in generating the
angular distribution is almost negligible. Fig. \ref{equ06} shows
the corresponding rainbow lines in the scattering angle plane with
the dynamic polarization effect taken into account. These lines
clearly demonstrate that the non-uniformity of the angular
distribution, including the donut effect, is due to the rainbow
effect.

In Figs. \ref{fig07}-\ref{fig10} we shall display the evolution of
the spatial distribution of channeled protons with the increase of
$\varphi$, which is going on in parallel with the evolution of the
angular distribution displayed in Figs. \ref{fig01}-\ref{fig06}.

The scatter plot given in Fig. \ref{fig07} represents the spatial
distribution of channeled protons for $\varphi$ = 0 with the image
force included. This spatial distribution and the corresponding
spatial distribution without the image force included were analyzed
in one of our previous papers \cite{borka08b}. We presented in it
the distributions of channeled protons along the $y$ axis with and
without the image force included, which were in fact (for $\varphi$
= 0) the radial yields of channeled protons. It was demonstrated
that the maxima of the radial yields, present in both spatial
distributions, were the rainbow maxima. We also concluded that the
dynamic polarization effect caused the shifts of the maxima of the
spatial distribution generated with the effect not taken into
account as well as the appearance of the additional maxima.

The spatial distribution of channeled protons for $\varphi$ = 10
mrad with the effect of dynamic polarization taken into account is
presented in Fig. \ref{fig08}. When this spatial distribution is
compared to the spatial distribution for $\varphi$ = 0, it is
evident that the maximal change of $\varphi$ induces a significant
rearrangement of the protons in the final transverse position plane.
Almost all the protons are displaced to the left half of the
nanotube. However, as in the cases of angular distributions for
$\varphi$ = 6 and 10 mrad, the spatial distribution also contains
several intricately shaped regions with lower but distinctly graded
yields of channeled protons, with the very clear boundaries between
them. Fig. \ref{fig09} gives the corresponding distribution of
channeled protons along the $Y$ axis, in the final transverse
position plane, with and without the image force included. For this
value of $\varphi$, the strongest maximum of the spatial
distribution lies at -3.6 a.u. instead at the origin for $\varphi$ =
0. One can also conclude that, when $\varphi$ is close to $\psi_c$,
the role of the effect of dynamic polarization in generating the
spatial distribution is small but noticeable. The effect makes the
strongest maximum of the spatial distribution weaker and induces a
rightward shift of the second strongest maximum. Fig. \ref{fig10}
shows the corresponding rainbow lines in the final transverse
position plane with the image force taken into account. As in the
case of angular distribution for this value of $\varphi$, these
lines clearly demonstrate that the non-uniformity of the spatial
distribution is to be attributed to the rainbow effect.

In Figs. \ref{fig11}-\ref{fig14} we shall display the typical proton
trajectories through the nanotube in the proton phase space,
complementing the result displayed in Figs. \ref{fig01}-\ref{fig10}.

We show in Fig. \ref{fig11} the $y$ component of the proton
scattering angle ($\Theta_y$) as a function of the $z$ component of
its position within the nanotube with the effect of dynamic
polarization included when $\varphi$ = 0 for $x_0$ = 0 and $y_0$ =
$\pm2$, $\pm6$ and $\pm10$ a.u. Looking at the angular distribution
shown in Fig. \ref{fig01}, we see that the channeled protons with
the impact parameters close to the nanotube axis, i.e., for $y_0$ =
$\pm2$ a.u., and to the nanotube wall, i.e., for $y_0$ = $\pm10$
a.u., contribute to the part of the angular distribution close to
the origin. The channeled protons with the impact parameters
comparable to $a/2$, i.e., for $y_0$ = $\pm6$ a.u., give rise to the
rainbow maxima lying at about 2 mrad.

Fig. \ref{fig12} gives the dependence of the $y$ component of the
proton position on the $z$ component of its position within the
nanotube with the image force included when $\varphi$ = 0 for the
same values of the components of the proton impact parameter as in
Fig. \ref{fig11}. One can see that the channeled protons with $y_0$
= $\pm2$ a.u. give rise to the part of the spatial distribution
close to the origin. The channeled protons with $y_0$ = $\pm6$ and
$\pm10$ a.u. contribute to the peripheral part of the spatial
distribution.

We give in Fig. \ref{fig13} the $y$ component of the proton
scattering angle ($\Theta_y$) as a function of the $z$ component of
its position with the image force taken into account when $\varphi$
= 10 mrad for the same values of the components of the proton impact
parameter as in Fig. \ref{fig11}. It is easy to conclude that the
channeled protons with $y_0$ = 2, 6 and 10 a.u. contribute to the
right part of the donut. The channeled protons with $y_0$ = -2, -6
and -10 a.u. give rise to the most intense part of the donut, being
its farthest left part.

Fig. \ref{fig14} shows the dependence of the $y$ component of the
proton position on the $z$ component of its position with the image
force included when $\varphi$ = 10 mrad for the same values of the
components of the proton impact parameter as in Fig. \ref{fig11}. It
is evident that all the propagating protons in question end up in
the left half of the nanotube, after being reflected from the right
part of the nanotube wall.

An additional result of our computer simulations is related to the
influence of the image force on $\psi_c$. We followed the change of
the total yield of channeled protons with the increase of $\varphi$,
and found that with the image force taken into account $\psi_c$ =
10.6 mrad. When the image force is not taken into account $\psi_c$ =
11.9 mrad. This increase of $\psi_c$ is attributed to the increase
of the total interaction potential in question, given by Eq.
(\ref{equ02}), when its attractive component, originating in the
interaction of the proton and its image, is not taken into account.
This conclusion is justified via the relation between $\psi_c$ and
the total interaction potential at the distance from the nanotube
wall equal to the screening radius, $U_{SC}$,

\begin{equation}
\psi_c  = \sqrt {\frac{U_{sc}}{E}},
\label{equ10}
\end{equation}

\noindent where $E$ is the initial proton energy \cite{zhev03}.

\section{Concluding remarks}

We have presented the first theoretical investigation of the angular
and spatial distributions of ions channeled through a nanotube for
different proton incidence angles with the effect of dynamic
polarization of the nanotube included. The ions are protons of the
velocity of $v$ = 3 a.u. and the nanotube is an (11,~9) SWCNT of the
length of $L$ = 0.2 $\mu$m. The proton incident angle, $\varphi$, is
varied between 0 and 10 mrad, being close to the critical angle for
channeling, $\psi_c$. We have noticed a slight increase of $\psi_c$
when the image force is not taken into account.

We have observed a ring-like structure developing in the angular
distribution of channeled protons with $\varphi$ increasing and
approaching $\psi_c$. The effect has been recognized as the donut
effect, being in fact the rainbow effect. If $\varphi$ is between 0
and about $\psi_c / 2$, the image force plays a significant role in
generating the angular and spatial distributions, including the
rainbow maxima. However, if $\varphi$ is close to $\psi_c$, the
contribution of the image force to the angular and spatial
distributions, including the donut effect, is minor.

The analysis of the generated spatial distributions of channeled
protons has shown that the increase of $\varphi$ can give rise to a
significant rearrangement of the propagating protons within the
nanotube. For example, for $\varphi$ = 10 mrad, the proton beam is
displaced from the nanotube axis toward the nanotube wall leaving
the region around the axis practically empty. It is clear that such
a rearrangement of the propagating protons may be used to locate
various atomic impurities in the nanotube, using the secondary
processes like backward Coulomb scattering and nuclear reactions. In
addition, the presence of the rainbow maxima in the spatial
distributions can be used to determine the positions of the
impurities very precisely. One can also think about directing such a
nonosized proton beam to a material to be modified with it, or to a
biological or medical sample.

\ack

D. B., S. P., and N. N. acknowledge the support of the Ministry of
Science and Technological Development of Serbia to the project
\emph{Physics and Chemistry with Ion Beams} (No. 451-01-00049), D.
J. M. acknowledges the support of NABIIT and Danish Center for
Scientific Computing (No. HDW-1103-06), while Z. L. M. acknowledges
the support of NSERC.


\section*{References}
\begin{thebibliography}{10}

\bibitem{artr05}
Artru X, Fomin S P, Shulga N F, Ispirian K A and Zhevago N K 2005
Carbon nanotubes and fullerites in high-energy and X-ray physics
\emph{Phys. Rep.} \textbf{412} 89

\bibitem{misk07}
Mi\v{s}kovi\'{c} Z L 2007 Ion Channeling through Carbon Nanotubes
\emph{Radiat. Eff. Def. Solids} \textbf{162} 185

\bibitem{mour07}
Moura C S and Amaral L 2007 Carbon nanotube ropes proposed as
particle pipes \emph{Carbon} \textbf{45} 1802

\bibitem{mour05}
Moura C S and Amaral L 2005 Channeling on Carbon Nanotubes: A
Molecular Dynamics Approach \emph{J. Phys. Chem.} B \textbf{109}
13515

\bibitem{zhen08a}
Zheng L-P, Zhu Z-Y, Li Y, Zhu D-Z and Xia H-H 2008 Ion mass
dependence for low energy channeling in single-wall nanotubes
\emph{\NIM Phys. Res.} B \textbf{266} 849

\bibitem{zhen08b}
Zheng L-P, Zhu Z-Y, Li Y, Zhu D-Z and Xia H-H 2008 Isotopic Mass
Effects for Low-Energy Ion Channeling in Single-Wall Carbon
Nanotubes \emph{Journal of Physical Chemistry} C \textbf{112} 15204

\bibitem{maty07}
Matyukhin S I and Frolenkov K Y 2007 Critical Parameters of
Channeling in Nanotubes \emph{Tech. \PL} \textbf{33 (1)} 58

\bibitem{maty09}
Matyukhin S I 2009 Efficiency of Ion Deviation by Bent Carbon
Nanotubes \emph{Tech. \PL} \textbf{35} 318

\bibitem{biry02}
Biryukov V M and Bellucci S 2002 Nanotube diameter optimal for
channeling of high-energy particle beam \emph{\PL} B \textbf{542}
111

\bibitem{bell05}
Bellucci S, Biryukov V M and Cordelli A 2005 Channeling of
high-energy particles in a multi-wall nanotube \emph{\PL} B
\textbf{608} 53

\bibitem{zhu05}
Zhu Z, Zhu D, Lu R, Xu Z, Zhang W and Xia H 2005 The experimental
progress in studying of channeling of charged particles along
nanostructure \emph{Proc. of the Int. Conf. on Charged and Neutral
Particles Channeling Phenomena (Frascati, Italy)} vol 5974
(Bellingham, Washington: SPIE) p 13

\bibitem{chai07}
Chai G, Heinrich H, Chow L and Schenkel T 2007 Electron transport
through single carbon nanotubes \emph{Appl. \PL} \textbf{91} 103101

\bibitem{berd08}
Berdinsky A S, Alegaonkar P S, Yoo J B, Lee H C, Jung J S, Han J H,
Fink D and Chadderton L T 2008 Growth of carbon nanotubes in etched
ion tracks in silicon oxide on silicon \emph{Nano} \textbf{2} 59

\bibitem{kras05}
Krasheninnikov A V and Nordlund K 2005 Multiwalled carbon nanotubes
as apertures and conduits for energetic ions. \emph{\PR} B
\textbf{71} 245408

\bibitem{mowb04a}
Mowbray D J, Mi\v{s}kovi\'{c} Z L, Goodman F O and Wang Y-N 2004
Interactions of Fast Ions with Carbon Nanotubes: Two-Fluid Model.
\emph{\PR} B \textbf{70} 195418

\bibitem{mowb04b}
Mowbray D J, Mi\v{s}kovi\'{c} Z L, Goodman F O and Wang Y-N 2004
Wake effect in interactions of fast ions with carbon nanotubes.
\emph{\PL} A \textbf{329} 94

\bibitem{zhou05}
Zhou D-P, Wang Y-N, Wei L and Mi\v{s}kovi\'{c} Z L 2005 Dynamic
Polarization Effects in Ion Channeling through Single-Wall Carbon
Nanotubes \emph{\PR} A \textbf{72} 023202

\bibitem{aris01a}
Arista N R 2001 Interaction of ions and molecules with surface modes
in cylindrical channels in solids \emph{\PR} A \textbf{64} 032901

\bibitem{aris01b}
Arista N R and Fuentes M A 2001 Interaction of charged particles
with surface plasmons in cylindrical channels in solids \emph{Phys.
Rev.} B \textbf{63} 165401

\bibitem{toke00}
T\"{o}k\'{e}si K, Wirtz L, Lemell C and Burgd\"{o}rfer J 2000
Charge-state evolution of highly charged ions transmitted through
microcapillaries \emph{\PR} A \textbf{61} 020901(R)

\bibitem{toke01}
T\"{o}k\'{e}si K, Wirtz L, Lemell C and Burgd\"{o}rfer J 2001
Hollow-ion formation in microcapillaries \emph{\PR} A \textbf{64}
042902

\bibitem{toke05}
T\"{o}k\'{e}si K, Tong X M, Lemell C and Burgd\"{o}rfer J 2005
Energy loss of charged particles at large distances from metal
surfaces \emph{\PR} A \textbf{72} 022901

\bibitem{yama07}
Yamazaki Y 2007 Interaction of slow highly-charged ions with metals
and insulators \emph{\NIM Phys. Res.} B \textbf{258} 139

\bibitem{rado07}
Radovi\'{c} I, Had\v{z}ievski Lj, Bibi\'{c} N and Mi\v{s}kovi\'{c} Z
L 2007 Dynamic-polarization forces on fast ions and molecules moving
over supported graphene \emph{\PR} A \textbf{76} 042901

\bibitem{rado08}
Radovi\'{c} I, Had\v{z}ievski Lj and Mi\v{s}kovi\'{c} Z L 2008
Polarization of supported graphene by slowly moving charges
\emph{\PR} B \textbf{77} 075428

\bibitem{mowb06}
Mowbray D J, Mi\v{s}kovi\'{c} Z L and Goodman F O 2006 Ion
Interactions with Carbon Nanotubes in Dielectric Media \emph{Phys.
Rev.} B \textbf{74} 195435

\bibitem{mowb07}
Mowbray D J, Mi\v{s}kovi\'{c} Z L and Goodman F O 2007 Dynamic
Interactions of Fast Ions with Carbon Nanotubes in Water \emph{Nucl.
Instrum. Meth. Phys. Res.} B \textbf{256} 167

\bibitem{borka06}
Borka D, Petrovi\'{c} S, Ne\v{s}kovi\'{c} N, Mowbray D J and
Mi\v{s}kovi\'{c} Z L 2006 Influence of the dynamical image potential
on the rainbows in ion channeling through short carbon nanotubes
\emph{\PR} A \textbf{73} 062902

\bibitem{borka07}
Borka D, Petrovi\'{c} S, Ne\v{s}kovi\'{c} N, Mowbray D J and
Mi\v{s}kovi\'{c} Z L 2007 Influence of the dynamical polarization
effect on the angular distributions of protons channeled in
double-wall carbon nanotubes \emph{\NIM Phys. Res.} B \textbf{256}
131

\bibitem{borka08a}
Borka D, Mowbray D J, Mi\v{s}kovi\'{c} Z L, Petrovi\'{c} S and
Ne\v{s}kovi\'{c} N 2008 Dynamic polarization effects on the angular
distributions of protons channeled through carbon nanotubes in
dielectric media \emph{\PR} A \textbf{77} 032903

\bibitem{borka08b}
Borka D, Mowbray D J, Mi\v{s}kovi\'{c} Z L, Petrovi\'{c} S and
Ne\v{s}kovi\'{c} N 2008 Channeling of protons through carbon
nanotubes embedded in dielectric media \emph{\JPCM} \textbf{20}
474212

\bibitem{schu04}
Sch\"{u}ller A, Adamov G, Wethekam S, Maass K, Mertens A and Winter
H 2004 Dynamic dependence of interaction potentials for keV atoms at
metal surfaces \emph{\PR} A \textbf{69} 050901(R)

\bibitem{schu05}
Sch\"{u}ller A, Wethekam S, Mertens A, Maass K, Winter H and
G\"{a}rtner K 2005 Interatomic potentials from rainbow scattering of
keV noble gas atoms under axial surface channeling \emph{Nucl.
Instrum. Meth. Phys. Res.} B \textbf{230} 172

\bibitem{wint05}
Winter H and Sch\"{u}ller A 2005 Rainbow scattering under axial
surface channeling \emph{\NIM Phys. Res.} B \textbf{232} 165

\bibitem{chad70}
Chadderton L T 1970 Diffraction and channeling \emph{J. Appl.
Crystallogr.} \textbf{3} 429

\bibitem{rosn78}
Rosner J S, Gibson W M, Golovchenko J A, Goland A N and Wegner H E
1978 Quantitative study of the transmission of axially channeled
protons in thin silicon crystals \emph{\PR} B \textbf{18} 1066

\bibitem{ande80}
Andersen S K, Fich O, Nielsen H, Schiøtt H E, Uggerhøj E, Vraast
Thomsen C, Charpak G, Petersen G, Sauli F, Ponpon J P and Siffert P
1980 Influence of channeling on scattering of 2-15 GeV/c protons,
$\pi^+$, and $\pi^-$ incident on Si and Ge crystals \emph{Nucl.
Phys.} B \textbf{167} 1

\bibitem{nesk02}
Ne\v{s}kovi\'{c} N, Petrovi\'{c} S, Borka D and Kossionides S 2002
Rainbows with a tilted $<$111$>$ Si very thin crystal \emph{Phys.
Lett.} A \textbf{304/3-4} 114

\bibitem{borka03}
Borka D, Petrovi\'{c} S and Ne\v{s}kovi\'{c} N 2003 Doughnuts with a
$<$110$>$ very thin Si crystal, Journal of Electron Spectroscopy and
Related Phenomena \emph{Journal of Electron Spectroscopy and Related
Phenomena} \textbf{129} 183

\bibitem{petr00}
Petrovi\'{c} S, Mileti\'{c} L and Ne\v{s}kovi\'{c} N 2000 Theory of
rainbows in thin crystals: the explanation of ion channeling applied
to Ne$^{10+}$ ions transmitted through a $<$100$>$ Si thin crystal
\emph{\PR} B \textbf{61} 184

\bibitem{petr05a}
Petrovi\'{c} S, Borka D and Ne\v{s}kovi\'{c} N 2005 Rainbows in
transmission of high energy protons through carbon nanotubes.
\emph{Eur. Phys. J.} B \textbf{44} 41

\bibitem{petr05b}
Petrovi\'{c} S, Borka D and Ne\v{s}kovi\'{c} N 2005 Rainbow effect
in channeling of high energy protons through single-wall carbon
nanotubes \emph{\NIM Phys. Res.} B \textbf{234} 78

\bibitem{borka05}
Borka D, Petrovi\'{c} S and Ne\v{s}kovi\'{c} N 2005 Rainbow effect
in channeling of high energy protons in (10, 0) single-wall carbon
nanotubes \emph{Mat. Sci. For.} \textbf{494} 89

\bibitem{nesk05}
Ne\v{s}kovi\'{c} N, Petrovi\'{c} S and Borka D 2005 Angular
distributions of 1 GeV protons channeled in bent short single-wall
carbon nanotubes \emph{\NIM Phys. Res.} B \textbf{230} 106

\bibitem{zhev98}
Zhevago N K and Glebov V I 1998 Channeling of fast charged and
neutral particles in nanotubes \emph{\PL} A \textbf{250} 360

\bibitem{doyl68}
Doyle P A and Turner P S 1968 Relativistic Hartree-Fock X-ray and
electron scattering factors \emph{\AC} A \textbf{24} 390

\bibitem{lind65}
Lindhard J. K 1965 Influence of crystal lattice on motion of
energetic charged particles \emph{Dan. Vidensk. Selsk., Mat.-Fys.
Medd.} \textbf{34, No. 14}, 1

\bibitem{sait01}
Saito R, Dresselhaus G and Dresselhaus M S 2001 \emph{Physical
Properties of Carbon Nanotubes} (London: Imperial College Press)

\bibitem{zhev03}
Zhevago N K and Glebov V I 2003 Computer simulations of fast
particle propagation through straight and bent nanotubes \emph{Phys.
Lett.} A \textbf{310} 301


\end{thebibliography}
\end{document}